\documentclass{article}
\usepackage{spconf,amsmath,graphicx}
\usepackage{multirow}
\usepackage{tipa}
\usepackage{subcaption}
\usepackage[hidelinks]{hyperref}
\usepackage{booktabs}
\usepackage{IEEEtrantools}
\usepackage[acronym]{glossaries}
\usepackage{amssymb}
\usepackage{todonotes}
\usepackage{caption}
\usepackage{subcaption}
\usepackage{float}
\usepackage{url}
\usepackage{diagbox}
\usepackage[symbol]{footmisc}
\title{Anonymizing Speech with Generative Adversarial Networks to Preserve Speaker Privacy}

\name{Sarina Meyer$^*$, Pascal Tilli$^*$, Pavel Denisov, Florian Lux, Julia Koch, Ngoc Thang Vu}
\address{Institute for Natural Language Processing (IMS), University of Stuttgart, Germany}
\copyrightnotice{978-1-6654-7189-3/22/\$31.00~\copyright2023 IEEE}
\begin{document}

\newacronym{gan}{GAN}{Generative Adversarial Network}
\newacronym{gans}{GANs}{Generative Adversarial Networks}
\newacronym{wgan}{WGAN}{Wasserstein Generative Adversarial Network}
\newacronym{wgans}{WGANs}{Wasserstein Generative Adversarial Networks}
\newacronym{wganqc}{WGAN-QC}{Wasserstein GAN with Quadratic Transport Cost}
\newacronym{otr}{OTR}{Optimal Transport Regularizer}
\newacronym{cnns}{CNNs}{Convolutional Neural Networks}
\newacronym{tdnn}{TDNN}{Time Delay Neural Network}
\newacronym{js}{JS}{Jensen-Shannon}
\newacronym{tsne}{t-SNE}{t-Distributed Stochastic Neighbor Embedding}
\newacronym{asr}{ASR}{speech recognition}
\newacronym{tts}{TTS}{text-to-speech}
\newacronym{wer}{WER}{Word Error Rate}
\newacronym{per}{PER}{Phone Error Rate}
\newacronym{eer}{EER}{Equal Error Rate}
\newacronym{gvd}{GVD}{Gain of Voice Distinctiveness}
\newacronym{mlp}{MLP}{Multilayer Perceptron}

%\ninept   % If I understand the styleguide correctly, it is allowed to use this option regardless of whether you want to do a 5 page journal submission or a 6 pages conference submission?
%
\newcommand\blfootnote[1]{%
  \begingroup
  \renewcommand\thefootnote{}\footnote{#1}%
  \addtocounter{footnote}{-1}%
  \endgroup
}

\maketitle
\begin{abstract}
In order to protect the privacy of speech data, speaker anonymization aims for hiding the identity of a speaker by changing the voice in speech recordings. This typically comes with a privacy-utility trade-off between protection of individuals and usability of the data for downstream applications. One of the challenges in this context is to create non-existent voices that sound as natural as possible.

In this work, we propose to tackle this issue by generating speaker embeddings using a generative adversarial network with Wasserstein distance as cost function. By incorporating these artificial embeddings into a speech-to-text-to-speech pipeline, we outperform previous approaches in terms of privacy and utility. According to standard objective metrics and human evaluation, our approach generates intelligible and content-preserving yet privacy-protecting versions of the original recordings.
\end{abstract}
\blfootnote{Copyright 2023 IEEE. Published in the 2022 IEEE Spoken Language Technology Workshop (SLT) (SLT 2022), scheduled for 19-22 January 2023 in Doha, Qatar. Personal use of this material is permitted. However, permission to reprint/republish this material for advertising or promotional purposes or for creating new collective works for resale or redistribution to servers or lists, or to reuse any copyrighted component of this work in other works, must be obtained from the IEEE. Contact: Manager, Copyrights and Permissions / IEEE Service Center / 445 Hoes Lane / P.O. Box 1331 / Piscataway, NJ 08855-1331, USA. Telephone: + Intl. 908-562-3966.}

\section{Introduction}
Within the last decades, speaker verification and identification systems have been improved up to a performance that allows applications in various settings, from access permissions to forensics.
However, the better such systems perform, the higher the risk of being abused in a harmful way.
In many applications, the transmission of speech recordings from local devices to cloud services does not require to sustain the identity of the speaker.
As a result, the demand of techniques that obfuscate the speakers' identity has risen.
Speaker anonymization describes the task of manipulating speech data such that the speaker identity becomes unrecognizable but every other information contained in the original speech - like the actual linguistic content - remains intact.

Until recently, speaker anonymization has gained little attention within the speech processing community and the existing work lacked consistency in definitions and metrics, making them difficult to compare. 
In order to change this, the Voice Privacy challenge was introduced in 2020 \cite{tomashenko2020introducing}. 
%initiative was founded with a first challenge at Interspeech 2020 \cite{tomashenko2020introducing}. 
%This challenge did not only increase the popularity of speaker anonymization but also provides a framework including conditions, data, metrics and baseline systems for future publications to base on.

The two challenge baselines proposed by \cite{srivastava_2020_design, patino21speaker} represent two different research directions which most anonymization approaches follow: (i) signal processing techniques and (ii) machine learning-based systems modifying speaker embeddings for speech synthesis.
%Speaker embeddings are speech representations learned to distinguish different speakers, e.g., for tasks like speaker diarization and speaker recognition. 
%They are used by the majority of voice privacy techniques which fall mostly into the machine learning category. 
%While they generally outperform signal processing methods in terms of objective privacy and utility metrics, they produce speech that is perceived as less natural and intelligible \cite{tomashenko2020voiceprivacy}. 
While the latter is used by the majority of approaches and generally outperform signal processing methods in terms of objective privacy and utility metrics, they produce speech that is perceived as less natural and intelligible \cite{tomashenko2020voiceprivacy}. 
One reason for this is that the machine-learning based approaches rely on complex and error-prone models as pipeline components, like \gls{asr} and \gls{tts} systems.
Another issue, as pointed out by \cite{turner2020speaker}, is that they tend to generate an anonymized speaker space that follows a different, thus unnatural, distribution than the original data.
This paper proposes to use \gls{gans} to fix this issue.

\gls{gans} have been shown to be a powerful framework to train generative models in an unsupervised manner \cite{gan_goodfellow, sagan}.
Over the years, many different methods have been proposed, including \gls{wgans} that offer increased optimization properties \cite{wgan, improved_wgan, wgan_qc}. Furthermore, the  generative nature of \gls{gans} draws interest to applications such as security and privacy that focus on leveraging the \emph{realness} of artificial data without having to worry about specific information that might be encoded in the data \cite{gans_sec_privacy_survey}. Thus, GANs seem to be a suitable choice to generate realistic but unknown speaker embeddings for speaker anonymization tasks.

\renewcommand{\thefootnote}{\fnsymbol{footnote}}
\footnotetext[1]{Equal contribution.}
\setcounter{footnote}{0}
\renewcommand{\thefootnote}{\arabic{footnote}}

In this paper, we present a novel approach\footnote{Our code and demo audios are publicly available at \url{https://github.com/DigitalPhonetics/speaker-anonymization}.} to sample non-existent voices for speaker anonymization by creating an artificial speaker embedding space using a \gls{wganqc} \cite{wgan_qc}.%Wasserstein Generative Adversarial Networks with Quadratic Transport Cost \cite{wgan_qc}.
We show that the distribution of the generated embedding space is similar to the one of the original recordings, effectively eliminating the distribution problem stated above. 
By using these artificial embeddings in a speech-to-text-to-speech anonymization framework, our approach achieves better speech recognition results and equally high privacy and voice distinctiveness scores as previous techniques, significantly outperforming the baseline of the Voice Privacy Challenge 2020. 
Based on a user study, we confirm high perceived privacy, linkability and speech quality of the anonymized utterances.

\section{Related Work}

\subsection{Speaker Anonymization} \label{subsec:rl_anon}
Signal processing in voice privacy comprises
formant-shifting techniques based on McAdams coefficients \cite{patino21speaker}, frequency warping \cite{qian_2017_voicemask, qian_2021_speech}, or a sequence of several signal processing steps \cite{kai_2021_lightweight, kai_2022_lightweight}, and by specifically modifying pitch \cite{tavi_2022_improving} and speech rate \cite{dubagunta_2022_adjustable}. These approaches typically have the advantage of being independent of training data or huge parameter sets, making them small and fast in execution.

However, the more popular research direction consists of techniques using machine learning models to modify the speaker information in speech. They usually first extract information regarding the pitch, linguistic content and speaker from the original recording, modify the speaker part, and use all to synthesize a new, anonymized version of the speech. This structure is used in the primary baseline of the challenge \cite{srivastava_2020_design} and in subsequent work \cite{turner2020speaker, espinoza2020speaker, champion2020speaker, champion2021astudy, mawalim2022speaker, miao_2022_language-independent}. \cite{champion2021astudy} found it beneficial to also modify the pitch before synthesis, and \cite{champion2020speaker} propose to include a masking of speaker-related information already in the training of the ASR system used to extract the content of the speech. \cite{meyer_2022_speaker} found that it is not necessary to include pitch to produce anonymized speech which in turn reduces the leakage of speaker information that is contained in pitch. A different approach is taken by \cite{yoo_2020_speaker} which apply an auto-encoder method to spectrograms in which the speaker information is modified during decoding.

\subsection{Speaker Embeddings} \label{subsec:rl_embeddings}
The main differences between the machine-learning based approaches lie in the modification of the extracted speaker information, typically in form of speaker embeddings, to transform them towards a new speaker.
Most approaches use x-vectors \cite{snyder_2018_x-vectors} as embedding model \cite{srivastava_2020_design, turner2020speaker, espinoza2020speaker, champion2020speaker, mawalim2022speaker, noe_2021_adversarial} but recently also ECAPA-TDNN vectors \cite{DBLP:conf/interspeech/DesplanquesTD20} are applied for speaker anonymization \cite{miao_2022_language-independent, meyer_2022_speaker}. \cite{meyer_2022_speaker} find that x-vector and ECAPA-TDNN actually complement each other in terms of speaker information and thus use the concatenation of both. In this work, we follow \cite{meyer_2022_speaker} by applying both vector types in combination.

In order to anonymize the speaker of an utterance, a new speaker embedding needs to be created that corresponds to an ideally non-existent speaker with a clearly different voice than the original speaker. The primary baseline of the challenge solves this by sampling from an external pool of speakers multiple x-vectors that are most distant from the x-vector of the original speaker, and generate a new artificial embedding as the average over the sampled ones. However, this has been found to lead to a different distribution of the anonymized embeddings as compared to the original ones. \cite{turner2020speaker} therefore propose to fit a Gaussian Mixture Model (GMM) to the dimension reduced x-vector space and use it to sample new vectors. They find this to improve the privacy of the anonymized speech in different attack scenarios. Another approach by \cite{mawalim2022speaker} selects a pseudo target speaker via clustering of the external speaker pool and transforms the original x-vector towards this target by modifying the singular values of the vector. 

As pointed out by \cite{turner2020speaker} and \cite{meyer_2022_speaker}, a mismatch in the distributions of original and anonymized speaker vectors is not ideal, and privacy can be improved if this difference is decreased. In this work, we therefore tackle this issue by using \gls{gans}.

\subsection{Generative Adversarial Networks}
Several speaker anonymization systems use adversarial auto-encoders or \gls{gans} to improve voice privacy. However, they use them to either disentangle multiple private attributes like sex and accent in addition to identity in order to increase control in anonymization \cite{espinoza2020speaker, prajapati_2021_voice}  -- and even to change an attribute like sex but keeping the original identity \cite{noe_2021_adversarial} --, or to disentangle the speaker information from the speech without using explicit speaker embeddings \cite{yoo_2020_speaker}. Although \gls{gans} have been applied to speaker embeddings for data augmentation \cite{wang_2020_data}, to our knowledge, no work has yet attempted to use this for speaker anonymization by mimicking the properties of the original speaker vector space in order to sample artificial yet natural-like embeddings.

\section{Proposed System with GAN-generated Speaker Embeddings}
\subsection{Speaker Vector based Anonymization System} 
\label{sec:system}
\begin{figure}[ht]
    %\centering
    \resizebox{0.49\textwidth}{!}{
    \includegraphics[scale=0.38]{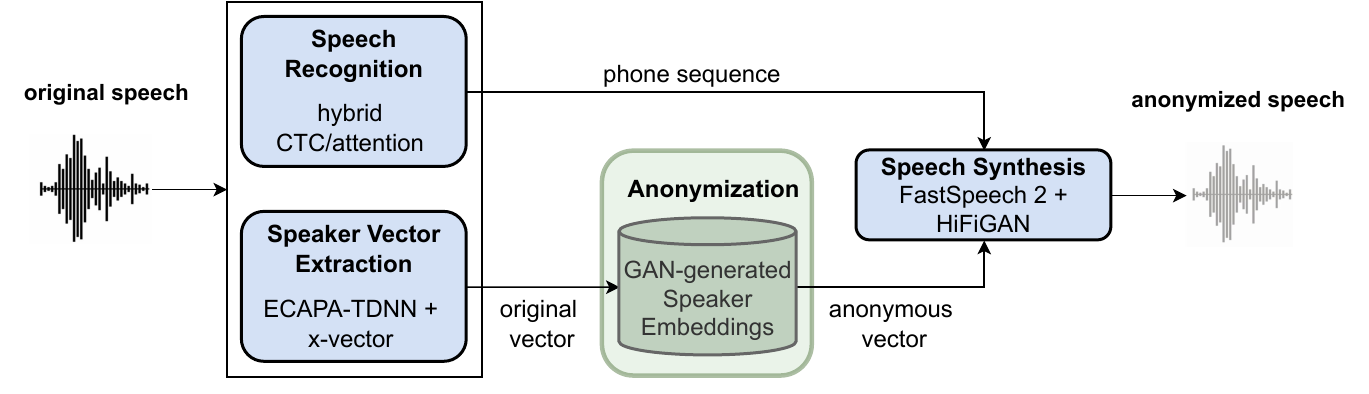}
    }
    \caption{Architecture of the proposed system with the GAN-based speaker anonymization.} % as novel component as compared to \cite{meyer_2022_speaker}.}
    \label{fig:architecture}
\end{figure}
Our approach follows the typical structure of machine learning-based voice anonymization with three phases as shown in Figure \ref{fig:architecture}: (i) an information extraction phase, (ii) an embedding modification phase, and (iii) a synthesis phase. %This structure with the corresponding modules is shown in Figure \ref{fig:architecture}. 
Since \cite{meyer_2022_speaker} have proposed a framework with high-performing \gls{asr} and \gls{tts} components, we use their system and only exchange the anonymization module (marked in green). This is an \gls{asr} model based on the hybrid CTC/attention architecture \cite{watanabe2017hybrid} with a Conformer as encoder \cite{gulati2020conformer} and a Transformer decoder. It is implemented in the ESPnet2 toolkit \cite{watanabe20212020}. The output of this model are phone sequences.
%The output of this module is not text, as typical for speech recognition, but phone sequences. 
For deriving speaker embeddings from speech, we use the x-vector and ECAPA-TDNN extractors provided by SpeechBrain \cite{speechbrain} and concatenate both vectors for creating one single embedding per utterance. Finally, the \gls{tts} component, implemented in the IMS Toucan toolkit \cite{lux2021toucan}, uses a FastSpeech 2 model \cite{ren2020fastspeech} for synthesizing the incoming phone sequence into spectrograms and a HiFiGAN  vocoder \cite{kong2020hifi} to translate them into waveforms. The synthesis is conditioned on real speaker embeddings to produce corresponding voices for different speakers. We use the same models as \cite{meyer_2022_speaker} but further optimized the hyperparameters of the \gls{asr} system for lower phone error rate. %: trained with CTC loss weight 0.6 instead of 0.3 and gradient accumulation over 8 steps instead of 4, averaged 10 best checkpoints instead of the best one, and decoded with CTC score weight 0.2 instead of 0.4. The Phone Error Rate values on the LibriSpeech dev/test clean and VCTK dev/test subsets are 6.5\%/6.4\% and 5.0\%/9.4\% instead of 7.4\%/7.2\% and 6.6\%/10.7\% in \cite{meyer_2022_speaker}.% for a better handling of the untypical phone output.

\subsection{Generating Artificial Speakers}
We use a \gls{wganqc} to generate artificial speaker embeddings in order to anonymize the original ones.
As in the vanilla GAN approach, a \gls{wganqc} consists of a generator and a discriminator.
The discriminator is often called \emph{critic} since it is not trained to classify between \emph{real} or \emph{fake} data but to decrease the distance between real and fake distributions.
Our target (real) data distribution $\mathbb{P}_r$ consists of $704$-dimensional speaker embeddings -- one per utterance in the training data -- which are a concatenation of $192d$ ECAPA-TDNN vectors and $512d$ x-vectors.

The generator receives a random vector $z$ sampled from a standard normal distribution $\mathcal{N}(0,1)$ as input and outputs a vector the same shape as our real speaker embeddings.
The critic is trained to compute the quadratic Wasserstein distance \cite{wgan_qc} w.r.t. the generated data samples and the original speaker embeddings.
Additionally, \cite{wgan_qc} compute the quadratic transport cost to further improve the convergence of \gls{wganqc}. We refer interested readers to \cite{wgan_qc} for any details about the architecture and training process of the model.

For our generator and critic models, we use ResNets \cite{resnet} based on \gls{cnns} as proposed by \cite{wgan_qc}.
Furthermore, we experiment with \gls{mlp} as generator and critic instead of ResNets since we operate on an arguably simpler domain than the typical task of generating images. 
%Furthermore, we experiment with \gls{mlp} as generator and critic instead of ResNets since we operate on an arguably simpler domain than generating images (which \gls{gans} are mostly evaluated in \cite{gan_goodfellow, wgan, improved_wgan, wgan_qc}). 
%Detailed discussion can be found in Section \ref{subsec:results_hyper}.

During anonymization, the generator samples an artificial embedding and compares it to the speaker embedding from the original recording. If the cosine distance between both vectors is above 0.3, the generated embedding is kept as new speaker representation. Otherwise, the sampling process is repeated until the condition is met. The selection of an artificial target speaker is performed once per input speaker and dataset to keep the same anonymous voice for all utterances of a speaker within a session.

%The successes of \gls{gans} are often demonstrated qualitatively on image generation tasks where it intuitively makes sense to provide samples created by the generator.
%In our case, visualizing speaker embeddings tends to be more difficult.
%We therefore present plots using the dimensionality reduction technique \gls{tsne} \cite{tsne} in this work to show how well the distribution of the generated samples matches the one of real data.

\section{Experimental Setup} \label{sec:exp_setup}

\subsection{Data}
We restrict all data to the guidelines of the Voice Privacy Challenge 2020 \cite{tomashenko2020introducing} in order to allow for comparability to other approaches. Thus, we use the x-vector and ECAPA-TDNN embedding extractors trained by SpeechBrain \cite{speechbrain} on the 2,800 h speaker verification corpus VoxCeleb 1 and 2 \cite{Nagrani19, Nagrani17, Chung18b}.
The speech recognition model was trained on LibriTTS \cite{zen2019libritts} spanning 600 h in total. For the TTS system, however, we use only a subset of the data, clean-100, that contains 100 h of clean speech. 

The evaluation data as given in the challenge consists of development and test splits of the LibriSpeech \cite{panayotov2015librispeech} and the VCTK \cite{yamagishi2019vctk} corpora. VCTK is divided into two sets, \textit{common} with the same sentences for all speakers and \textit{different} in which the uttered sentences differ between speakers. Each split is divided into enrollment and trial data. The enrollment data serves as reference data for the speaker verification attacker (see Section \ref{subsec:eval_metrics}) whereas the trial data is the one from which the speaker identity should be concealed.

\subsection{Objective Evaluation} 
\label{subsec:eval_metrics}
Three objective metrics are applied to the models. They are computed using the evaluation models of the challenge. 
%They are computed within the evaluation suite of the challenge, using the evaluation models provided with it.

In order to assess the privacy strength of each approach, an automatic speaker verification (ASV) attacker is applied to the anonymized trial data and its performance measured as Equal Error Rate (\textbf{EER}). 
We aim for an EER of 50\%, denoting a random decision behavior by the attacker. 
Since the ASV model is only used for evaluation without giving it the possibility to change its strategy based on the correctness of its prediction, we additionally favor EER scores above 50\% because this means that the attacker makes more mistakes, not knowing that it should flip its predictions.
Two attack scenarios are tested: The \textit{ignorant} attacker has only access to the original enrollment data and tries to relate that to the anonymized trial data. On the other hand, the \textit{lazy-informed} attacker uses enrollment data that has been anonymized by the same technique as the trial data but with different target speakers. In that case, the ASV system is a stronger opponent because it can exploit any speaker-specific artifacts that are still remained after anonymization.
 
Two utility metrics are applied.
%to measure the speech information beyond speaker identity preserved in the anonymized data. 
One assesses the remaining linguistic content and intelligibility as Word Error Rate (\textbf{WER}), measured using \gls{asr}.
As a second metric, the Gain of Voice Distinctiveness (\textbf{GVD}, \cite{noe20speech}) is computed. It measures to which degree the distinctiveness between voices of different speakers is kept as compared to the non-anonymized data. 
A GVD score of zero denotes the same voice distinctiveness as in the original data, below zero a decrease in distinctiveness, and above zero an increase. Everything close to zero or above is desired in this task.

\subsection{Subjective Evaluation}
We verify our results in a human evaluation study,
mainly following the one conducted in the challenge.
Since anonymized speech should be distinctive between different speakers but staying consistent for each individual speaker within a session, we collect ratings on speaker \textbf{linkability} between two anonymized utterances. 
Users have to decide if the two utterances come from the \textit{same} or \textit{different} speakers but can also select the option \textit{unsure}. We randomly selected 20 pairs of utterances with half coming from the same original speaker for both recordings and the rest from different ones.

We further evaluate speaker \textbf{verifiability} in order to test whether anonymized speech can still be linked to the original speaker. The participants listen to audio pairs consisting of an enrollment sample from an original speaker and a trial sample which may be either from the \textit{same or a different original speaker} and may be \textit{anonymized or not}, resulting in four combinations to evaluate. We ask the users to rate the similarity of the speakers of the two utterances as well as \textbf{naturalness} and \textbf{intelligibility} of the trial sample on a scale from 1 to 5. The study comprises 6 items for each scenario.

\subsection{Baselines}
We compare the performance of our approach to four baselines: (i) The original, non-anonymized data (\textit{original}), (ii) the results reported for the primary baseline of the Voice Privacy Challenge 2020 (\textit{BL VPC20}), (iii) to a re-implementation of the anonymization method of this primary baseline as described by \cite{meyer_2022_speaker} (\textit{pool}), and (iv), also used by \cite{meyer_2022_speaker}, a baseline with random speaker embeddings (\textit{random}). Since (iii) and (iv) use the pipeline described in this paper, they differ from the proposed system only in the way how speaker embeddings are modified. 
The main differences between (ii) and (iii) are (a) the use of different TTS and ASR models, (b) the use of phone information in (iii) instead of bottleneck features, (c) the use of pitch values in (ii), (d) the concatenation of ECAPA-TDNN and x-vector in (iii) but use of only x-vector in (ii), and (e) a normalization of the averaged embeddings in (iii) to avoid unnatural value ranges.

\subsection{WGAN Settings} \label{subsec:wgan_settings}
Due to a limited data size, we chose to drastically reduce the size of the ResNets from \cite{wgan_qc}.
The generator and critic in our model contain of $150,000$ parameters across three residual blocks.
We experiment with different dimensions for the input $z\sim\mathcal{N}(\mu,\,\sigma^{2})$ ($\mu=0,\,\sigma^{2}=1$) of the generator, $16$ and $64$.
For $\gamma$, which is mentioned to be tuned in a range of $[0.01, 1]$ by \cite{wgan_qc}, 
they achieved best results by using a values of $0.1$ and $1.0$.
Consequently, we decided to limit the hyperparameter search in Section \ref{subsec:results_hyper} to these two values of $\gamma$.
%These results can be found in Section \ref{subsec:results_hyper}.
Unless stated otherwise, the results in this paper where achieved with a GAN using 16-dimensional input noise and $\gamma = 1.0$.

\section{Results}
\label{sec:exp_results}

\subsection{Privacy}
The privacy results in form of EER for each attack scenario are given in Table \ref{tab:results_eer_oa} and Table \ref{tab:results_eer_aa}. For the ignorant scenario in which the attacker bases the prediction on original enrollment data, all models perform similarly well with EER scores close to 50\% or even above. In the lazy-informed scenario, on the other hand, the approaches using the architecture in Figure \ref{fig:architecture} clearly outperform the challenge baseline, and show an equally well performance as in the ignorant case. Regarding privacy, there are no significant differences between \textit{random}, \textit{pool} and \textit{GAN} which all use the same pipeline architecture. 

\begin{table}[th]
    \centering
    \resizebox{\linewidth}{!}{
    \begin{tabular}{c|cc|cc|cc}
        \toprule
        & \multicolumn{2}{c}{LibriSpeech} & \multicolumn{2}{|c}{VCTK-diff} & \multicolumn{2}{|c}{VCTK-comm}\\
        Model & F & M & F & M & F & M \\
        \midrule
        \textit{Original} & \textit{7.66} & \textit{1.11} & \textit{4.89} & \textit{2.07} & \textit{2.89} & \textit{1.13}\\
        \midrule
        BL VPC20 & 47.26 & \textbf{52.12} & 48.05 & \textbf{53.85} & 48.27 & 53.39\\
        Random & 49.27 & 48.55 & 58.49 & 48.68 & 50.00 & 49.44\\
        Pool & \textbf{55.11} & 50.78 & \textbf{58.59} & 52.30 & \textbf{52.31} & 50.28\\
        \midrule
        %GAN & 48.72 & \textbf{55.23} & 43.31 & 42.71 & 50.00 & 48.02\\
        GAN & 48.36 & 48.11 & 51.95 & 50.40 & 47.98 & \textbf{54.24}\\
        \bottomrule
    \end{tabular}
    }
    \caption{\textbf{EER} (in \%) as privacy metric in the \textbf{ignorant} scenario (original enrollment, anonymized trial data), for \textbf{F}emale and \textbf{M}ale separately. Higher is better.}
    \label{tab:results_eer_oa}
\end{table}

\begin{table}[th]
    \centering
    \resizebox{\linewidth}{!}{
    \begin{tabular}{c|cc|cc|cc}
        \toprule
        & \multicolumn{2}{c}{LibriSpeech} & \multicolumn{2}{|c}{VCTK-diff} & \multicolumn{2}{|c}{VCTK-comm}\\
        Model & F & M & F & M & F & M \\
        \midrule
        \textit{Original} & \textit{7.66} & \textit{1.11} & \textit{4.89} & \textit{2.07} & \textit{2.89} & \textit{1.13}\\
        \midrule
        BL VPC20 & 32.12 & 36.75 & 31.74 & 30.94 & 31.21 & 31.07\\
        Random & 43.61 & 49.22 & \textbf{54.94} & \textbf{52.93} & \textbf{51.73} & 48.59\\
        Pool & 45.07 & \textbf{51.22} & 52.16 & 49.02 & 49.71 & 43.50\\
        \midrule
        %GAN & \textbf{52.37} & 51.00 & \textbf{55.25} & \textbf{55.97} & \textbf{54.34} & \textbf{52.54}\\
        GAN & \textbf{53.83} & 46.10 & 49.43 & 52.47 & 47.98 & \textbf{48.87}\\
        \bottomrule
    \end{tabular}
    }
    \caption{\textbf{EER} (in \%) as privacy metric in the \textbf{lazy-informed} scenario (anonymized enrollment and trial data), for \textbf{F}emale and \textbf{M}ale separately. Higher is better.}
    \label{tab:results_eer_aa}
\end{table}

\subsection{Utility}
\begin{table}[h]
    \centering
    \small{
    \begin{tabular}{c|cc}
        \toprule
        Model  & LibriSpeech & VCTK \\
        \midrule
        \textit{Original} & 4.15 & 12.82\\
        \midrule
        BL VPC 20 & 6.73 & 15.23\\
        Random & 6.18 & 10.98\\
        Pool & 6.78 & 11.38\\
        \midrule
        %GAN & \textbf{5.95} & \textbf{10.03}\\
        GAN & \textbf{5.90} & \textbf{10.02}\\
        \bottomrule
    \end{tabular}
    }
    \caption{\textbf{WER} (in \%) as speech recognition utility metric.}
    \label{tab:results_wer}
\end{table}
The WER scores of the speech recognition evaluation are shown in Table \ref{tab:results_wer}. One strength of the pipeline is the use of high-performing \gls{asr} and \gls{tts} models which are optimized for recognizing and producing high quality speech. This advantage is visible in the results of the WER metric for which the challenge baseline is outperformed by all models on VCTK and almost all on LibriSpeech. On both datasets, our GAN-based approach achieves the best performance which is even better than for the original speech on VCTK. Since VCTK is generally a challenging corpus for speech recognition because it consists of recordings in different accents, this finding suggests that our \gls{asr} model is better performing than the one used for evaluation and that our anonymization method decreases the accent information in the speech, thus increasing the privacy level. This indicates that the anonymized speech produced by our system is not only suitable for downstream applications requiring the original speech content, but also contribute highly to the requirement of the voice privacy task in retaining the linguistic information. Out of the participants of the Voice Privacy Challenge 2020 as presented in \cite{tomashenko2020voiceprivacy}, only one group achieved a slightly lower WER of 5.8 than us and none came close to our results for VCTK \footnote{The best group of the challenge for VCTK reached a WER of 14.6.}.

\begin{table}[h]
    \centering
    \resizebox{\linewidth}{!}{
    \begin{tabular}{c|cc|cc|cc}
        \toprule
        & \multicolumn{2}{c}{LibriSpeech} & \multicolumn{2}{|c}{VCTK-diff} & \multicolumn{2}{|c}{VCTK-comm}\\
        Model  & F & M & F & M & F & M\\
        \midrule
        BL VPC20 & -10.09 & -8.95 & -10.55 & -11.58 & -9.65 & -10.39\\
        Random & -0.25 & -0.31 & -0.61 & -1.13 & 0.02 & -0.13\\
        Pool & -0.17 & \textbf{-0.04} & -0.06 & \textbf{-0.36} & -0.03 & \textbf{-0.05}\\
        \midrule
        %GAN & -0.24 & -0.25 & \textbf{-0.03} & -0.50 & \textbf{0.05} & -0.13\\
        GAN & \textbf{-0.06} & -0.15 & \textbf{0.18} & -0.41 & \textbf{0.07} & -0.16\\
        \bottomrule
    \end{tabular}
    }
    \caption{\textbf{GVD} as voice distinctiveness utility metric. Closer to zero or above is better.}
    \label{tab:results_gvd}
\end{table}

A different condition of the task is to create distinctive voices that could be distinguished in a conversation. This is measured by the GVD metric and shown in Table \ref{tab:results_gvd}. Again, the methods using the same pipeline achieve scores close to zero, suggesting a similar voice distinctiveness as in the original data, whereas the challenge baseline produces less distinctive voices. The GAN approach performs equally well in this metric as the pool method that is inspired by the baseline.

\subsection{Comparison to Related Work}
So far, we compared our system only with approaches that follow a clearly different anonymization technique which were not developed with the goal to create natural-like speaker embeddings. In order to evaluate our method in relation to previous work with similar focus or methodology, Table \ref{tab:results_rw} displays a comparison to two participants of the Voice Privacy Challenge 2020 as reported in their results \cite{tomashenko2020voiceprivacy}. \cite{turner2020speaker} identify the distribution mismatch of the baseline anonymization and thus present an approach applying a GMM on the dimension reduced pool x-vector space and sampling new x-vectors from this GMM. This technique is similar to ours in that anonymized speaker embeddings are drawn from a distribution similar to the original one. \cite{espinoza2020speaker} do not follow this objective but resemble our system in using an adversarial auto-encoder that has a similar functionality as a GAN. 

\begin{table}[ht]
    \centering
    \resizebox{\linewidth}{!}{
    \begin{tabular}{c|cc|cc|cc}
         \toprule
         & \multicolumn{2}{c}{EER} & \multicolumn{2}{|c}{GVD} & \multicolumn{2}{|c}{WER}\\
        Approach & Libri & VCTK & Libri & VCTK & Libri & VCTK \\
        \midrule
        \cite{turner2020speaker} & 40.89 & 37.65 & -12.14 & -13.79 & 7.1 & 15.6\\
        \cite{espinoza2020speaker} & 41.29 & 37.35 & -13.60 & -15.22 & 6.8 & 15.2\\
        % Our & \textbf{51.69} & \textbf{55.61} & \textbf{-0.25} & \textbf{-0.27} & \textbf{5.95} & \textbf{10.03}\\
         Ours & \textbf{49.97} & \textbf{50.95} & \textbf{-0.11} & \textbf{-0.12} & \textbf{5.9} & \textbf{10.02}\\
         \bottomrule
    \end{tabular}
    }
    \caption{Comparison to related approaches, averaged over female and male.
    %. For VCTK, only the \textit{diff} set is shown, and the results for both datasets are averaged over female and male. The EER is given for the lazy-informed scenario.
    }
    \label{tab:results_rw}
\end{table}

For simplicity, the scores in Table \ref{tab:results_rw} are averaged over female and male, only VCTK-diff is shown for VCTK and only the lazy-informed scenario for EER. Compared to our system, both approaches exhibit similar results for all metrics as the challenge baseline shown in Tables \ref{tab:results_eer_aa} to \ref{tab:results_gvd}, and are all outperformed significantly by our proposed method.

\section{Analysis}
\begin{figure*}[h!tb]
\begin{subfigure}{0.24\textwidth}
    \includegraphics[width=\textwidth]{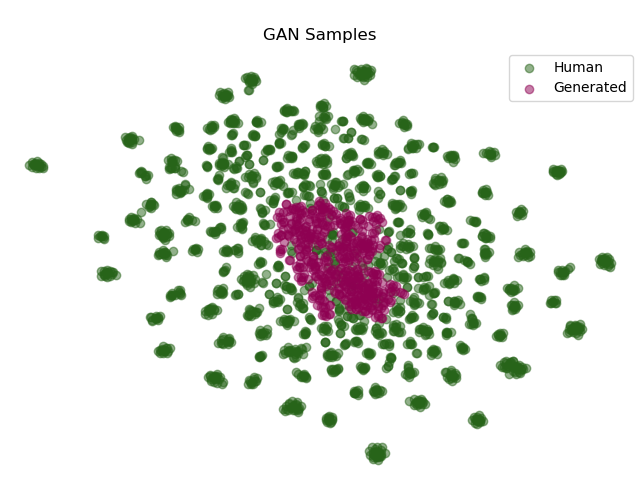}
    \caption{After $\sim5$k iterations.}
    \label{fig:first}
\end{subfigure}
\begin{subfigure}{0.24\textwidth}
    \includegraphics[width=\textwidth]{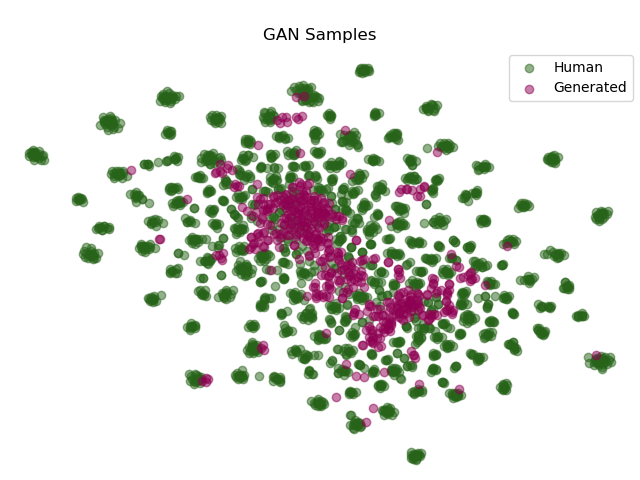}
    \caption{After $\sim35$k iterations.}
    \label{fig:second}
\end{subfigure}
\begin{subfigure}{0.24\textwidth}
    \includegraphics[width=\textwidth]{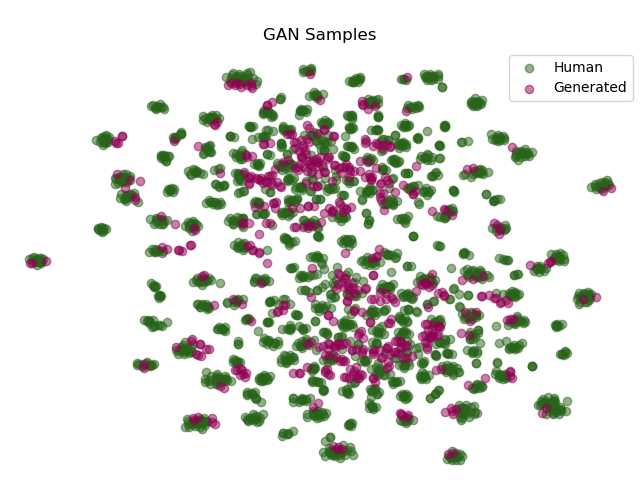}
    \caption{After $\sim135$k iterations.}
    \label{fig:third}
\end{subfigure}
\begin{subfigure}{0.24\textwidth}
    \includegraphics[width=\textwidth]{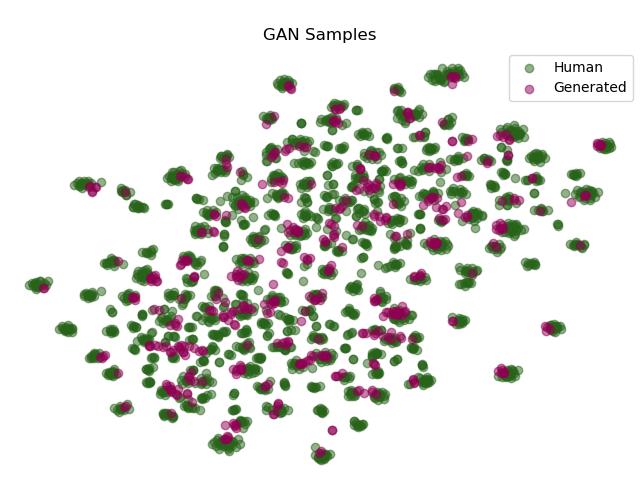}
    \caption{After $>200$k iterations.}
    \label{fig:fourth}
\end{subfigure}
\caption{Speaker embeddings of real (green) and artificial (purple) voices as produced by the WGAN after different numbers of training iterations. Dimensionality reduction was performed with \gls{tsne}.}
\label{fig:gan_samples}
\end{figure*}

\subsection{User Study} \label{sec:user_study}
%We present the results of our human evaluation study in the following. 
%For our linkability study, 
\begin{table}[]
    \centering
    \small{
    \begin{tabular}{c|r|r|r}
    \toprule
     \diagbox{true}{answer} & same & different & unsure \\
    \midrule
      same speaker &  \textbf{85.63} & 04.37 & 10.0\\
      different speaker & 06.87 & \textbf{85.63} & 07.50 \\
    \bottomrule
    \end{tabular}
    }
    \caption{Aggregated ratings scores in \% for \textbf{linkability} between anonymized audios. Rows represent the true relation between the audios and columns the user answers.}
    \label{tab:linkability}
\end{table}

For the linkability questions in our user study, 
we received responses from 16 participants, giving us 160 answers for same and different speaker scenario each.  The results are displayed in Table \ref{tab:linkability}.
With 85.63\% in both cases, the vast majority of answers are correct, indicating that human listeners can distinguish between anonymized speakers. We also observe that users would rather admit to be unsure than to give a wrong answer which further reduces the misidentified samples. This shows that our system is in fact able to maintain consistent voices for a speaker and at the same time generates distinctive ones for different speakers.

The mean opinion scores (MOS) for the resulting subjective metrics are given in Table \ref{tab:verifiability}. With ratings from 18 participants, we obtain a total of 108 ratings for each score and configuration. Regarding verifiability, unsurprisingly same speakers in enrollment and trial sample could be easily identified when the trial sample was not anonymized. In contrast, when the trial data was anonymized, the scores are close to the similarity ratings for different original speakers, even in the case where both utterances originally come from the same speaker. In other words, after anonymization speakers were perceived as different to almost the same extent as real speakers differ from each other, proving the success of our anonymization method. 

We observe a moderate gap in naturalness of 1.2 points on average on the 1-5 Likert scale  between anonymized, i.e. synthesized samples compared to non-anonymized human recordings. Nevertheless, there is only a minor decrease in intelligibility. So, although the speech produced by our system might be perceived as synthetic to some extent, the linguistic content is still preserved.

\begin{table}[]
    \centering
    \resizebox{\linewidth}{!}{
    \begin{tabular}{ll|r|r|r|r|r|r}
    \toprule
speaker   & anon & \multicolumn{2}{c|}{verifiability} & \multicolumn{2}{c|}{naturalness} & \multicolumn{2}{c}{intelligibility} \\

          &      & MOS           & $\sigma$          & MOS          & $\sigma$         & MOS            & $\sigma$           \\
\midrule
same      & no   & 4.13          & $\pm1.22$         & 4.67         & $\pm0.49$        & 4.74           & $\pm0.44$          \\
same      & yes  & 1.58          & $\pm0.97$         & 3.44         & $\pm1.00$        & 4.48           & $\pm0.77$          \\
different & no   & 1.24          & $\pm0.65$         & 4.33         & $\pm0.95$        & 4.5            & $\pm0.89$          \\
different & yes  & 1.67          & $\pm1.08$         & 3.09         & $\pm1.23$        & 4.18           & $\pm0.93$   \\      
        \bottomrule
    \end{tabular}
    }
    %\caption{MOS scores on a scale from 1-5 for speaker \textbf{verifiabilty}, \textbf{naturalness} and \textbf{intelligibility} for 18 participants on 24 items. Users were instructed to rate similarity between an enrollment sample from an original speaker and a trial sample either from the same or different speaker either anonymized or not. Naturalness and intelligibility scores refer to the trial sample.}
    \caption{MOS scores on a scale from 1-5 for speaker \textbf{verifiability} between trial and enrollment utterances, and \textbf{naturalness} and \textbf{intelligibility} of trial samples.}
    \label{tab:verifiability}
\end{table}

%\begin{table}[]
%    \centering
%    \resizebox{\linewidth}{!}{
%    \begin{tabular}{ll|r|r|r}
%    \toprule
%        speaker & anon & verifiability & naturalness & intelligibility \\
%         \midrule
%        same & no & 4.13 & 4.67 & 4.74\\
%        same & yes & 1.58 & 3.44 & 4.48 \\
%        different & no & 1.24 & 4.33 & 4.5\\
%        different & yes & 1.67 & 3.09 & 4.18 \\
%        \bottomrule
%    \end{tabular}
%    }
%    \caption{MOS scores on a scale from 1-5 for speaker \textbf{verifiability} between trial and enrollment utterances, and \textbf{naturalness} and \textbf{intelligibility} of trial samples.}
%    \label{tab:verifiability}
%\end{table}

\subsection{Naturalness of WGAN Embeddings}
\iffalse
\begin{figure*}[h!tb]
\begin{subfigure}{0.24\textwidth}
    \includegraphics[width=\textwidth]{figures/12-07-2022-16-48-15_pca+tsne_step_5190.png}
    \caption{After $\sim5$k iterations.}
    \label{fig:first}
\end{subfigure}
\begin{subfigure}{0.24\textwidth}
    \includegraphics[width=\textwidth]{figures/12-07-2022-16-48-15_pca+tsne_step_36330.png}
    \caption{After $\sim35$k iterations.}
    \label{fig:second}
\end{subfigure}
\begin{subfigure}{0.24\textwidth}
    \includegraphics[width=\textwidth]{figures/12-07-2022-16-48-15_pca+tsne_step_134940.png}
    \caption{After $\sim135$k iterations.}
    \label{fig:third}
\end{subfigure}
\begin{subfigure}{0.24\textwidth}
    \includegraphics[width=\textwidth]{figures/12-07-2022-16-48-15_pca+tsne_step_342540.png}
    \caption{After $>200$k iterations.}
    \label{fig:fourth}
\end{subfigure}
\caption{Speaker embeddings of real (green) and artificial (purple) voices as produced by the WGAN after different numbers of training iterations. Dimensionality reduction was performed with \gls{tsne}.}
\label{fig:gan_samples}
\end{figure*}
\fi

To analyze the naturalness of our generated speaker embeddings, we project them along real speaker embeddings into a two-dimensional space using \gls{tsne} to visualize the results.
They are shown in Figure \ref{fig:gan_samples}.
It is visible that after $5$k iterations, the generator is already capable of producing embeddings which are in distribution of the human embeddings but do not cover the whole variance.
With increasing training iterations, \gls{tsne} becomes unable to distinguish between generated samples or real samples, resulting in the generated data being spread all over the plot.

\subsection{Hyperparameters and Architecture of WGAN} \label{subsec:results_hyper}
\iffalse
\begin{figure}[h]
    \centering
    \resizebox{\linewidth}{!}{
    \begin{subfigure}{0.24\textwidth}
        \includegraphics[width=\textwidth]{figures/MLP-20-07-2022-17-28-52_pca+tsne_step_202410.png}
        \caption{MLP}
        \label{fig:200k_mlp}
    \end{subfigure}
    \begin{subfigure}{0.24\textwidth}
        \includegraphics[width=\textwidth]{figures/12-07-2022-16-48-15_pca+tsne_step_342540.png}
        \caption{ResNet}
        \label{fig:200k_resnet}
    \end{subfigure} 
    }
    \caption{Speaker embeddings of real (green) and artificial (purple) voices produced by \gls{wgans} of different architecture after more than 200k iterations. Dimensionality reduction was performed with \gls{tsne}.}
    \label{fig:resnet_vs_mlp}
\end{figure}
\fi

We tested the impact of the different settings mentioned in Section \ref{subsec:wgan_settings} on the embedding generation and resulting WER. Specifically, we evaluated the cases of (i) setting $\gamma$ to 0.1, (ii) using 64-dimensional random noise as input to the generator, and (iii) exchanging the ResNet model by a four layer \gls{mlp} that matches the number of trainable parameters. We found that (i) and (ii) slightly increased the WER, and that (i) slows down the convergence of the WGAN. Overall, both hyperparameters do not seem to affect the quality of the system much. This, however, is not the case for the model architecture: the use of \gls{mlp} in (iii) prevents the model to converge, leading to generated embeddings that can be easily distinguished from original ones. %, as shown in Figure \ref{fig:resnet_vs_mlp}. 
This suggests that the claim by \cite{wgan} stating that \gls{mlp} is unsuitable for training \gls{gans} is not only true for the image domain but also for speaker embeddings.

\section{Conclusions} 
\label{sec:conclusion}
In this paper, we presented a novel method for using Generative Adversarial Networks in a speaker anonymization framework in order to generate artificial speaker embeddings for new voices. By using Wasserstein distance and quadratic transport cost during training, we enforce the distribution of the generated embeddings to be similar of the one of speaker embeddings corresponding to real speakers. Including them into a speech-to-text-to-speech pipeline with high quality speech recognition and synthesis components leads to anonymized speech that preserves privacy, voice distinctiveness and linguistic content to a high degree. We show that the speaker embeddings generated by the network lead to more intelligibility and thus better speech recognition scores than other speaker embedding anonymization methods. By conducting a user study, we confirmed that our approach produces speech of high privacy and intelligibility.

% We can add this section later when the paper gets accepted
\section*{Acknowledgments}
Funded by Deutsche Forschungsgemeinschaft (DFG, German Research Foundation) under Germany's Excellence Strategy - EXC 2075 – 390740016. We acknowledge the support by the Stuttgart Center for Simulation Science (SimTech).

%\newpage
%\renewcommand{\bibfont}{\tiny}
\bibliographystyle{IEEEbib}
\bibliography{refs}

\end{document}